\documentclass[12pt]{article}

\usepackage{amsmath}
\usepackage{amssymb}
\usepackage[dvips]{graphicx}
\usepackage{setspace}

\newcommand{\ud}{\mathrm{d}}
\newcommand{\K}{{\sf{K}}}
\newcommand{\tsnr}{\text{\footnotesize{SNR}}}
\newcommand{\ssnr}{\text{\scriptsize{SNR}}}

\newtheorem{prop:optphase}{Proposition}
\newtheorem{prop:Kuhn-Tucker}[prop:optphase]{Proposition}
\newtheorem{prop:pnKuhn-Tucker}[prop:optphase]{Proposition}
\newtheorem{prop:discrete}[prop:optphase]{Proposition}
\newtheorem{prop:QAGC}[prop:optphase]{Proposition}
\newtheorem{theo:Rician}{Theorem}
\newtheorem{theo:peakRician}[theo:Rician]{Theorem}
\newtheorem{theo:pnRician}[theo:Rician]{Theorem}
\newtheorem{lemma:bessel}{Lemma}
\newtheorem{theo:Rician2}[theo:Rician]{Theorem}
\newtheorem{prop:twomass}[prop:optphase]{Proposition}
\newtheorem{prop:derivatives}[prop:optphase]{Proposition}
\newtheorem{cor:ebnozero}{Corollary}
\newtheorem{cor:widebandslope}[cor:ebnozero]{Corollary}
\newtheorem{conj:klessthenbeta}{Conjecture}
\newtheorem{prop:peakpower}[prop:optphase]{Proposition}
\newtheorem{prop:secondorderoptimal}[prop:optphase]{Proposition}
\newtheorem{def:BPSK}{Definition}
\newtheorem{def:QPSK}[def:BPSK]{Definition}
\newtheorem{prop:BPSK}[prop:optphase]{Proposition}
\newtheorem{prop:QPSK}[prop:optphase]{Proposition}
\newtheorem{cor:BPSKQPSKwidebandslope}[cor:ebnozero]{Corollary}
\newtheorem{lemma:kurtosis}{Lemma}

\begin{document}

\title{The Noncoherent Rician Fading
Channel -- Part I : Structure of the Capacity-Achieving Input
\footnote{This research was supported by the U.S. Army Research
Laboratory under contract DAAD 19-01-2-0011. The material in this
paper was presented in part at the Fortieth Annual Allerton
Conference on Communication, Control, and Computing, Monticello,
IL, Oct., 2002 and the Canadian Workshop on Information Theory,
Waterloo, Ontario, May 18-21, 2003. }}
\author{Mustafa Cenk Gursoy \and H. Vincent Poor \and Sergio Verd\'u}

\date{Dept. of Electrical
Engineering \\Princeton University \\[3.8pt] Princeton, NJ 08544}
\maketitle

\thispagestyle{empty}
\begin{spacing}{1.5}
%\begin{spacing}{1.1}
\begin{abstract}
Transmission of information over a discrete-time memoryless Rician
fading channel is considered where neither the receiver nor the
transmitter knows the fading coefficients. First the structure of
the capacity-achieving input signals is investigated when the
input is constrained to have limited peakedness by imposing either
a fourth moment or a peak constraint. When the input is subject to
second and fourth moment limitations, it is shown that the
capacity-achieving input amplitude distribution is discrete with a
finite number of mass points in the low-power regime. A similar
discrete structure for the optimal amplitude is proven over the
entire {\ssnr} range when there is only a peak power constraint.
The Rician fading with phase-noise channel model, where there is
phase uncertainty in the
specular component, is analyzed. % introduced.                                 %%%%%%%%%%%%%%%%%%SV
For this model it is shown that, with only an average power
constraint, the capacity-achieving input amplitude is discrete
with a finite number of levels. For the classical average power
limited Rician fading channel, it is proven that the optimal input
amplitude distribution has bounded support.
\\

\end{abstract}
\end{spacing}

\newpage
\begin{spacing}{1.6}
%\begin{spacing}{1.1}
\section{Introduction} \label{sec:intro}
%\begin{spacing}{1.1}
\setcounter{page}{1} Recently, the information theoretic analysis
of fading channels is receiving much attention. This interest is
motivated by the rapid advances in wireless technology and the
need to use scarce resources such as bandwidth and power as
efficiently as possible under severe fading conditions. Providing
the ultimate performance, information theoretic measures such as
capacity, spectral efficiency and error exponents can be used as
benchmarks to which we can compare the performance of practical
communication systems. Furthermore with the recent discovery of
codes that operate very close to the Shannon capacity, information
theoretic limits have gained practical relevance. Although the
capacity and other information-theoretic measures of fading
channels were investigated in as early as the 1960's
(\cite{Kennedy}, \cite{Richters}), it is only recently that many
interesting fading channel models have been considered under
various practically related input and channel constraints.

A significant amount of effort has been expended to study fading
channel models where side information about the fading is
available at either the receiver or the transmitter or both (see
\cite{Goldsmith Varaiya}, \cite{Caire Shamai}, \cite{Medard
Goldsmith}, \cite{Medard}). However, under fast fading conditions
noncoherent communications, where neither party knows the fading,
often becomes the only available alternative. Richters
\cite{Richters} considered the problem of communicating over an
average power limited discrete-time memoryless Rayleigh fading
channel without any channel side information. He conjectured that
the capacity-achieving amplitude distribution is discrete with a
finite number of mass points. Recently, Abou-Faycal \emph{et al.}
\cite{Abou} gave a rigorous proof of Richters' conjecture. This
result shows that when the fading is known by neither the
transmitter nor the receiver, the optimal amplitude distribution
has a notably different character than that of unfaded Gaussian
channels. A similar discrete structure for the optimal input was
also shown in \cite{Shamai} for the pulse amplitude modulated
direct detection photon channel when average and peak power
limitations are imposed on the intensity of a photon emitting
source. Katz and Shamai \cite{Katz Shamai} considered the
noncoherent AWGN channel and proved that the optimal input
amplitude is discrete with an infinite number of mass points.
Lapidoth \cite{Lapidoth} recently analyzed the effects of phase
noise over the AWGN channel, characterizing the high-{\tsnr}
asymptotics of the channel capacity for a general class of phase
noise distributions with memory. An extensive study of the
capacity of multi-antenna fading channels at high {\tsnr} was
conducted in \cite{Lapidoth Moser}.

Kennedy \cite{Kennedy} showed that the infinite bandwidth capacity
of fading multipath channels is the same as that of the unfaded
Gaussian channel. Although any set of orthogonal signals achieves
this capacity for the unfaded Gaussian channel, orthogonal signals
that are peaky both in time and frequency are needed in the
presence of fading \cite[Sec. 8.6]{Gallager}. Indeed, for a
general class of fading channels, Verd\'u \cite{Verdu} has
recently shown that if there are no constraints other than average
power, {\em flash signaling}, a class of unbounded peak-to-average
ratio inputs defined in \cite{Verdu}, is necessary to achieve the
capacity as $\textrm{\footnotesize{SNR}}
\rightarrow 0$ when the channel realization is unknown at the receiver.      %%%%%%%%%%%%%%%%%%SV
Flash signaling can be practically employed in systems where
sudden discharge of energy (e.g., using capacitors) is allowed,
thus sidestepping the use of RF amplifiers. However, these peaky
signals are not feasible in communication systems subject to
strict peak-to-average ratio requirements. Furthermore, in some
systems CDMA-type white signals which spread their energy over the
available bandwidth are used because of their anti-jamming and low
probability of intercept capabilities. Hence, it is of interest to
investigate the effect upon the capacity of imposing peakedness
constraints, especially in the low-power regime. M\'edard and
Gallager \cite{Medard Gallager} considered noncoherent broadband
fading channels with no specular component and limited the
peakedness of the input signals by imposing a fourth moment
constraint. Then they showed that such a constraint forces the
mutual information to zero inversely with increasing bandwidth.
Since CDMA-type signals spread their energy over the available
bandwidth, they satisfy the above fourth moment constraint.
Therefore, M\'edard and Gallager conclude that CDMA-type white
signals cannot efficiently utilize fading
multipath channels at extremely large bandwidth. Other results on this theme        %%%%%%%%%%%%%%%SV
are obtained by Telatar and Tse \cite{Telatar Tse} and Subramanian
and Hajek \cite{Subramanian Hajek}.

In this paper, we consider the noncoherent discrete-time
memoryless Rician fading channel and study the capacity and the
optimal input structure. In a companion paper [Part II], we
further investigate the spectral-efficiency/bit-energy tradeoff in
the low-power regime. The organization of the paper is as follows.
In Section \ref{sec:model}, we introduce the Rician fading channel
model. In Section \ref{sec:Rician}, we characterize the structure
of the capacity achieving input distribution in the low-power
regime when the channel input is constrained to have limited
peakedness. In Section \ref{sec:pnRician}, the average power
limited Rician fading with phase-noise channel, where there is
phase uncertainty in the specular component, is introduced and the
structure of the capacity-achieving input is investigated.
Numerical results are given in Section \ref{sec:Numerical}, while
Section \ref{sec:conclusion} contains our conclusions.

\section{Channel Model} \label{sec:model}

In this paper, we consider the following discrete-time memoryless
Rician fading channel model %\vspace{-0.5cm}
\begin{eqnarray}
y_i &=& mx_i + a_ix_i + n_i \label{eq:model}
\end{eqnarray}
where $\{a_i\}$ and $\{n_i\}$ are sequences of independent and
identically distributed (i.i.d.) circular zero mean complex
Gaussian random variables, independent of each other and of the
input, with variances $E\{|a_i|^2\} = \gamma^2$ and $E\{|n_i|^2\}
= N_0$, $m$ is a deterministic complex constant, $x_i$ is the
complex channel input and $y_i$ is the complex channel output.
$\{a_i\}$ and $\{n_i\}$ represent sequences of fading coefficients
and background noise
samples, respectively.     %%%%%%%%%%%%%%%%%%%%SV

The Rician fading channel model is particularly appropriate when
there is a direct line-of-sight (LOS) component in addition to the
faded component arising from multipath propagation. Moreover, the
Rician model includes both the unfaded Gaussian channel and the
Rayleigh fading channel as two special cases. Hence, results
obtained for this model provide a unifying perspective.

In the channel model (\ref{eq:model}), fading is assumed to be
flat and hence has a multiplicative effect on the channel input.
This is a valid assumption if the delay spread of the channel is
much smaller than the symbol duration. Moreover, frequency
selective fading channels can often be decomposed into parallel
non-interacting flat fading subchannels using orthogonal
multicarrier techniques. Note also that the fading coefficients
assume independent realizations at every symbol period. Under such
fast fading conditions, reliable estimation of the fading
coefficients may be quite difficult because of the short duration
between independent fades. Therefore, we consider the noncoherent
scenario where neither the receiver nor the transmitter knows the
fading coefficients $\{a_i\}$.

%%%%%%%%%%%%%%%%%%%%%%%%%%%%%%%%%%%%%%%%%%%%%%%%%%%%%%%%%%%%%%%%%%%%%%%%%%%%%%%%%%%%%%%%%
%%%%%%%%%%%%%%%%%%%%%%%%%%%%%%%%%%%%%%%%%%%%%%%%%%%%%%%%%%%%%%%%%%%%%%%%%%%%%%%%%%%%%%%%%
%%%%%%%%%%%%%%%%%%%%%%%%%%%%%%%%%%%%%%%%%%%%%%%%%%%%%%%%%%%%%%%%%%%%%%%%%%%%%%%%%%%%%%%%%
%%%%%%%%%%%%%%%%%%%%%%%%%%%%%%%%%%%%%%%%%%%%%%%%%%%%%%%%%%%%%%%%%%%%%%%%%%%%%%%%%%%%%%%%%

\section{Channel Capacity and Optimal Input Distribution} \label{sec:Rician}
\vspace{-0.2cm} In this section, we elaborate on the structure of
the capacity achieving input distribution for the Rician fading
channel when the input has limited peakedness which is achieved by
imposing a fourth moment or a peak limitation on the input
amplitude.

\subsection{Second and Fourth Moment Limited Input}
We first assume that the input amplitude is subject to second and
fourth moment constraints:
\begin{eqnarray}
E\left\{|x_i|^{2}\right\} &\leq& P_{av}  \qquad \,\,\,\forall i
\label{eq:constraint1}
\\ E\left\{|x_i|^{4}\right\} &\leq& \kappa P_{av}^{2} \qquad \forall i \label{eq:constraint2}
\end{eqnarray}
where $P_{av}$ is the average power constraint and $ 1< \kappa <
\infty$. When the average power constraint is active, the fourth
moment constraint is equivalent to limiting the kurtosis,
$\frac{E\left\{|x_i|^4\right\}}{\left(E\left\{|x_i|^2\right\}\right)^2}
\leq \kappa$, which is a measure of the peakedness of the input
signal. For the Rician channel model (\ref{eq:model}), the
capacity is the supremum of the input-output mutual information
over the set of all input distributions satisfying the constraints
(\ref{eq:constraint1}) and (\ref{eq:constraint2}) \footnote{Since
the channel is memoryless, without loss of generality we can drop
the index $i$.} \vspace{-0.57cm}
\begin{gather} \label{eq:cap}
C = \sup_{\substack{F_{x}(\cdot)\\ E\left\{|x|^{2}\right\} \leq
P_{av} \\ E\left\{|x|^{4}\right\}\leq \kappa P_{av}^2 }}
\int_{\mathbb{C}}\int_{\mathbb{C}} f_{y|x}(y|x) \log
\frac{f_{y|x}(y|x)}{f_y(y)} \, \, \ud y \, \, \ud F_x(x)
\end{gather}
where the conditional density of the output given the input,
\begin{gather} \label{eq:conditional}
f_{y|x}(y|x) = \frac{1}{\pi(\gamma^{2}|x|^{2}+N_0)} \exp\left(
-\frac{|y-mx|^{2}}{\gamma^{2}|x|^{2}+N_0} \right),
\end{gather}
is circular complex Gaussian. Moreover note that $f_y(y) =
\int_{\mathbb{C}} f_{y|x}(y|x)\, \ud F_x(x)$ is the marginal
output density, $F_x$ is the distribution function of the input
and $\mathbb{C}$ denotes the complex plane.

First we have the following preliminary result on the optimal
phase distribution.

\begin{prop:optphase} \label{prop:optphase}
For the Rician fading channel (\ref{eq:model}) %with $|m| > 0 $
and input constraints (\ref{eq:constraint1}) and
(\ref{eq:constraint2}), uniformly distributed phase that is
independent of the input amplitude is capacity-achieving
\footnote{The result holds in wider generality: the feasible set
defined by (\ref{eq:constraint1}) and (\ref{eq:constraint2}) can
be replaced by any set of constraints that are imposed only on the
input magnitude.}.
\end{prop:optphase}
\emph{Proof}: The result follows readily from the arguments in
\cite[Sec. IV.D.6]{Lapidoth Moser} where the optimality of
circularly symmetric input distributions is pointed out in a more
general setting. Assume that an input random variable $x$
generates a mutual information, $I_0$. Consider a new input $x_1 =
x e^{j\theta}$ where $\theta$ is independent of $x$ and uniformly
distributed on $(-\pi, \pi]$. Since the conditional distribution
has the property $f_{y|x}(e^{j\theta} y | e^{j\theta} x) =
f_{y|x}(y|x)$ for any $\theta$, it can be easily seen that mutual
information is invariant under deterministic rotations of the
input distribution and hence $I(x_1;y | \theta) = I_0$. By the
concavity of the mutual information over input distributions, we
have $I(x_1;y) \ge I_0$. Further note that input constraints
(\ref{eq:constraint1}) and (\ref{eq:constraint2}) are also
invariant to rotation. Therefore there is no loss in optimality in
considering input random variables that has uniformly distributed
phase independent of the amplitude. \hfill $\square$

Note that if $|m| = 0$, we have a Rayleigh fading channel where
the phase cannot %{\bf global edit change can not to cannot}                   %%%%%%%%%%%%%%%%%%SV
be used to convey information when the channel is unknown.
However, if there is a line-of-sight component, i.e., $|m|>0$,
phase can indeed carry information and by Proposition
\ref{prop:optphase}, the uniform distribution maximizes the
transmission rate.

With this characterization, we have reduced the optimization
problem (\ref{eq:cap}) to optimal selection of the distribution
function of the input amplitude, $F_{|x|}(\cdot)$, under the
constraints $E\{|x|^{2}\} \leq P_{av}$ and $E\{|x|^{4}\} \leq
\kappa P_{av}^2$. For the sake of simplification in the notation,
we introduce new random variables $R = \frac{1}{N_0}|y|^{2}$ and
$r = \frac{\gamma}{\sqrt{N_0}} |x|$. Assuming a uniform input
phase that is independent of the input magnitude,
%in (\ref{eq:polarmutual})
the mutual information in nats, after a straightforward
transformation, can be expressed as follows:
\begin{align}\label{eq:update2mutual}
I(F_r) \stackrel{\text{def}}{=} I(x;y) = -\int_{0}^{\infty} \!\!
f_R(R;F_r) \ln f_R(R;F_r) \, \ud R \! - \! \int_{0}^{\infty} \ln(1
+ r^{2}) \, \ud F_r(r) - 1
\end{align}
where $f_R(R;F_r) = \int_{0}^{\infty} g(R,r) \, \ud F_r(r)$ is the
density function of $R$ with the kernel given by $g(R,r) =
\frac{1}{1+r^{2}} \exp \left( -\frac{R+{\sf{K}}r^{2}}{1+r^{2}}
\right) I_0 \left( \frac{2\sqrt{{\sf{K}}}r \sqrt{R}}{1 + r^{2}}
\right)$. Furthermore, $F_r$ is the distribution function of $r$
and ${\sf{K}} = \frac{|m|^2}{\gamma^2}$ is the Rician factor. Now,
the capacity formula can be recast as follows
\begin{gather} \label{eq:capacity}
C(\alpha, \kappa, {\sf{K}}) = \sup_{ \substack{ F_r \\
E\{r^{2}\} \leq \alpha
\\ E\{r^4\} \leq \kappa \alpha^2}} I(F_r)
\end{gather}
depending on three parameters, namely, $\alpha = \gamma^2
\frac{P_{av}}{N_0}$, the normalized {\footnotesize{SNR}};
$\kappa$; and ${\sf{K}}$, the Rician factor. In \cite{gursoy}, the
existence of an optimal amplitude distribution achieving the
supremum in (\ref{eq:capacity}) is shown and the following
sufficient and necessary condition for an amplitude distribution
to be optimal is derived employing the techniques used in
\cite{Abou}.

%%%%%%%%%%%%%%%%%%%%%%%%%%%%%%%%%%%%%%%%%%%%%%%%%%%%%%%%%%%%%%%%%%%
%%%%%%%%%%%%%%%%%%%% KUHN-TUCKER CONDITION %%%%%%%%%%%%%%%%%%%%%%%%
%%%%%%%%%%%%%%%%%%%%%%%%%%%%%%%%%%%%%%%%%%%%%%%%%%%%%%%%%%%%%%%%%%%
\begin{prop:Kuhn-Tucker}(\underline{Kuhn-Tucker Condition}) \label{prop:ktc}
For the Rician channel (\ref{eq:model}) with input constraints
(\ref{eq:constraint1}) and (\ref{eq:constraint2}), $F_{0}$ is a
capacity achieving amplitude distribution if and only if there
exist $\lambda_1, \lambda_2 \geq 0$
%satisfying $\lambda_1 + 2\lambda_2k\alpha \geq0 $
such that the following is satisfied
\begin{eqnarray}
\int_0^{\infty}g(R,r) \ln f_R(R,F_0) \, \ud R + \ln(1+r^2)+
\lambda_{1}(r^{2}-\alpha)+\lambda_{2}(r^{4}-\kappa\alpha^{2})+ C+1
\geq 0 \quad \forall r \geq 0 \label{eq:ktc}
\end{eqnarray}
with equality if $r \in E_0$ where $E_0$ is the set of points of
increase\footnote{The set of points of increase of a distribution
function $F$ is $\{r: F(r-\epsilon) < F(r + \epsilon) \,\,\,
\forall \epsilon >0$\}.} of $F_0$.
\end{prop:Kuhn-Tucker}

%%%%%%%%%%%%%%%%%%%%%%%%%%%%%%%%%%%%%%%%%%%%%%%%%%%%%%%%%%%%%%%%%%%
%%%%%%%%%%%%%%%%%%%%%%  RICIAN CHANNEL %%%%%%%%%%%%%%%%%%%%%%%%%%%%
%%%%%%%%%%%%%%%%%%%%%%%%%%%%%%%%%%%%%%%%%%%%%%%%%%%%%%%%%%%%%%%%%%%
Note that in the above formulation, $\lambda_1$ and $\lambda_2$
are the Lagrange multipliers for the second and fourth moment
constraints respectively. Using Proposition \ref{prop:ktc}, we
have the following result on the optimal amplitude distribution.

\begin{theo:Rician} \label{theo:Rician}
For the Rician fading channel (\ref{eq:model}) with input
amplitude constraints (\ref{eq:constraint1}) and
(\ref{eq:constraint2}), if the fourth moment constraint
(\ref{eq:constraint2}) is active then the capacity-achieving input
amplitude distribution is discrete with a finite number of mass
points.
\end{theo:Rician}
\emph{Proof:} The result is shown by contradiction. The proof can
be summarized as follows:

i) We first contradict the assumption that the optimal
distribution has an infinite number of points of increase on a
bounded interval.

ii) Next, we contradict the assumption that the optimal
distribution has an infinite number of points of increase (mass
points) but only finitely many of them on any bounded interval.

iii) Ruling out the above assumptions leaves us with the only
possibility that the optimal input has a finite number of mass
points.

Assume $F_0$ is an optimal amplitude distribution. To prove the
theorem, we first find a lower bound on the left hand side (LHS)
of the Kuhn-Tucker condition (\ref{eq:ktc}). To that end, we first
bound $f_R$ as follows, \vspace{-0.5cm}
\begin{align}
f_R(R;F_0) &= \int_{0}^{\infty} g(R,r) \, \ud F_0(r) \nonumber \\
&\geq \int_{0}^{\infty} \frac{1}{1+r^{2}} \exp \left(
-\frac{R+{\sf{K}}r^{2}}{1+r^{2}} \right) \, \ud F_0(r) \\ &\geq
\exp(-R)\int_{0}^{\infty} \frac{1}{1+r^{2}} \exp \left(
-\frac{{\sf{K}}r^{2}}{1+r^{2}} \right) \, \ud F_0(r) \\ &=
D_{F_0}\exp(-R) \qquad \forall R \geq 0
\label{eq:lowerboundRician}
\end{align}
where $0<D_{F_0}\leq 1$ is a constant depending on $F_0$. The
first inequality is obtained from the fact that $I_0(x) \geq 1
\,\,\, \forall x \geq 0$ and the second inequality comes from
observing that $e^{-R/(1+r^2)} \geq e^{-R} \,\,\, \forall R,r \geq
0$. %Furthermore, since $g(R,r) \leq 1 \,\,\, \forall R,r \geq0$ ,
%$f_R(R,F_0) \leq 1 \,\,\, \forall R \geq 0$.
Using the lower bound (\ref{eq:lowerboundRician}), and noting that
$g(R,r)$ is a noncentral chi-square density function in $R$, we
have the following bound on the LHS of the Kuhn-Tucker condition
(\ref{eq:ktc}) \vspace{-0.3cm}
\begin{align}
\text{LHS} \geq \ln D_{F_0} - 1 -(1+{\sf{K}})r^2 +
\ln(1+r^2)+\lambda_{1}(r^{2}-\alpha)+\lambda_{2}(r^{4}-\kappa\alpha^{2})+1+C
\,\,\, \forall r\geq 0. \nonumber
\end{align}
If the fourth moment constraint is active, i.e., $\lambda_2 > 0$,
then for all $D_{F_0}>0$ and $\lambda_1 \ge 0$,  the above lower
bound diverges to infinity as $r \rightarrow \infty$. Now, we
establish contradiction in the following assumptions.
\\
i) Assume that the optimal distribution $F_0$ has an infinite
number of points of increase on a bounded interval. Note that this
assumption is satisfied by continuous distributions. First we
extend the LHS of the Kuhn-Tucker condition (\ref{eq:ktc}) to the
complex domain.
\begin{gather}\label{eq:ktccomplex}
\Phi(z) = \int_0^{\infty} g(R,z) \ln f_R(R,F_0) \, \ud R
+\ln(1+z^2)+
\lambda_{1}(z^{2}-\alpha)+\lambda_{2}(z^{4}-\kappa\alpha^{2})+ C+1
\end{gather}

\vspace{-0.3cm} \noindent $\!\!$where $z \in \mathbb{C}$. By the
Differentiation Lemma \cite[Ch. 12]{Lang}, it is easy to see that
$\Phi$ is analytic in the region where $\text{Re}\{1+z^2\}>0$
\cite[Appendix C]{gursoy}. This choice of region guarantees the
uniform convergence of the integral expression in
(\ref{eq:ktccomplex}). Note that in this region, by our earlier
assumption, $\Phi(z)= 0$ for an infinite number of points having a
limit point\footnote{The Bolzano-Weierstrass Theorem \cite{Rudin}
states that every bounded infinite set of real numbers has a limit
point.}. By the Identity Theorem\footnote{The Identity Theorem for
analytic functions \cite{Knopp} states that if two functions are
analytic in a region $\mathcal{R}$, and if they coincide in a
neighborhood, however small, of a point $z_0$ of $\mathcal{R}$, or
only along a path segment, however small, terminating in $z_0$, or
also only for an infinite number of distinct points with the limit
point $z_0$, then the two functions are equal everywhere in
$\mathcal{R}$.}, $\Phi(z) = 0$ in the region where
$\text{Re}\{1+z^2\}>0$. Hence the Kuhn-Tucker condition
(\ref{eq:ktc}) is satisfied with equality for all $r\geq 0$.
Clearly, this case is not possible from the above lower bound
which diverges to infinity as $r \rightarrow \infty$.
\\
ii) Next assume that the optimal distribution has an infinite
number of mass points but only finitely many of them on any
bounded interval. Then the LHS of the Kuhn-Tucker condition should
be equal to zero infinitely often as $r \rightarrow \infty$ which
is again not possible by the above diverging lower bound.
\\
Hence the optimal distribution must be discrete with a finite
number of mass points and the theorem follows. \hfill $\square$

The significance of Theorem \ref{theo:Rician} comes from the fact
that for the Rician fading channel, any fourth moment constraint
with a finite $\kappa$ will eventually be active for sufficiently
small {\footnotesize{SNR}} because, as observed in \cite[Sec.
V.E]{Verdu}, if there is no such constraint, the required value of
$\kappa$ grows without bound as $\tsnr \rightarrow 0$. Therefore,
Theorem \ref{theo:Rician} establishes the discrete nature of the
optimal input in the low-power regime. Furthermore, Theorem
\ref{theo:Rician} easily specializes to the Rayleigh and the
unfaded Gaussian channels. For the unfaded Gaussian channel, if
the fourth moment constraint is inactive, it is well known that a
Rayleigh distributed amplitude is optimal and it has kurtosis
$\kappa = 2$. Therefore the fourth moment constraint being active
(i.e., $1 < \kappa < 2$) is also a necessary condition for the
discrete nature in that case. In \cite{Abou}, the optimal
amplitude for the average-power-limited noncoherent Rayleigh
fading channel is shown to be discrete with a finite number of
levels over the entire {\tsnr} range. Theorem \ref{theo:Rician}
proves that this discrete character does not change when we have
an additional fourth moment constraint.

We note that the key property which leads to the proof of Theorem
\ref{theo:Rician} is that we have a moment constraint higher than
the second moment. Hence the result of Theorem \ref{theo:Rician}
holds in a more general setting where the fourth moment constraint
(\ref{eq:constraint2}) is replaced by a constraint in the
following form: $E\{|x|^{2+\delta}\} \leq M$ for some $\delta
> 0$ and $M < \infty$. In a related work, Palanki \cite{Palanki} has
independently shown the discrete character of the optimal input
for a general type of fading channels when only moment constraints
strictly higher than the second moment are imposed.

%%%%%%%%%%%%%%%%%%%%%%%%%%%%%%%%%%%%%%%%%%%%%%%%%%%%%%%%%%%%%%%%%%%%%%%%%%%%%%%%%%%%%%%%%
%%%%%%%%%%%%%%%%%%%%%%%%%%%%%%%%%%%%%%%%%%%%%%%%%%%%%%%%%%%%%%%%%%%%%%%%%%%%%%%%%%%%%%%%%
%%%%%%%%%%%%%%%%%%%%%%%%%%%%%%%%%%%%%%%%%%%%%%%%%%%%%%%%%%%%%%%%%%%%%%%%%%%%%%%%%%%%%%%%%
%%%%%%%%%%%%%%%%%%%%%%%%%%%%%%%%%%%%%%%%%%%%%%%%%%%%%%%%%%%%%%%%%%%%%%%%%%%%%%%%%%%%%%%%%

\subsection{Peak Power Limited Input}

In this section, we assume that the input amplitude is subject to
only a peak power limitation,
\begin{gather} \label{eq:peakconstraint}
|x_i|^2  \stackrel{\text{a.s.}}{\le} P \quad \forall i.
\end{gather}
Although being more stringent than the fourth moment limitation,
peak power constraint is more relevant in practical systems. For
instance, efficient use of battery power in portable radio units
and linear operation of RF amplifiers employed at the transmitter
require peak power limited communication schemes.

Since the peak constraint (\ref{eq:peakconstraint}) is invariant
to rotation of the input, optimality of uniform phase follows from
Proposition \ref{prop:optphase}. Hence, our primary focus is on
obtaining a characterization for the optimal amplitude
distribution. Existence of a capacity-achieving amplitude
distribution readily follows from the results on the second and
fourth moment limited case. Similarly, we define $R =
\frac{1}{N_0}|y|^2$ and $r = \frac{\gamma}{\sqrt{N_0}}|x|$.
Specializing (\ref{eq:ktc}), we easily obtain the following
sufficient and necessary condition for an amplitude distribution,
$F_0$, to be optimal over the peak-power limited Rician channel:
\begin{gather} \label{eq:ktcpeak}
\int_0^{\infty}g(R,r) \ln f_R(R,F_0) \, \ud R + \ln(1+r^2)+ C+1
\geq 0 \quad \forall r \in \left[0, \sqrt{\alpha} \right]
\end{gather}
with equality if $r \in E_0$ where $E_0$ is the set of points of
increase of $F_0$. Note that $\alpha = \gamma^2 \frac{P}{N_0}$.
Next, we state the main result on the optimal amplitude
distribution.

\begin{theo:peakRician} \label{theo:peakRician}
For the Rician fading channel (\ref{eq:model}) where the input is
subject to only a peak power constraint $|x|^2
\stackrel{\text{a.s.}}{\le} P$, the capacity-achieving amplitude
distribution is discrete with a finite number of mass points.
\end{theo:peakRician}
\emph{Proof}: Since the input is subject to a peak constraint, the
result in this case is established by contradicting the assumption
that the optimal input distribution has an infinite number of
points of increase on a bounded interval.

Assume $F_0$ is an optimal distribution. To prove the theorem, we
first find an upper bound on the left-hand-side (LHS) of
(\ref{eq:ktcpeak}). To achieve this goal, we bound
$f_R(\cdot,F_0)$ as follows.
\begin{align}
f_R(R;F_0) &= \int_{0}^{\sqrt{\alpha}} \frac{1}{1+r^{2}} \exp
\left( -\frac{R+{\sf{K}}r^{2}}{1+r^{2}} \right) I_0 \left(
\frac{2\sqrt{{\sf{K}}}r \sqrt{R}}{1 + r^{2}} \right) \, \ud F_0(r)
\nonumber
\\
&\le \exp\left( -\frac{R}{1+\alpha} + \sqrt{{\sf{K}}R} \right)
\int_{0}^{\sqrt{\alpha}} \frac{1}{1+r^{2}} \exp \left(
-\frac{{\sf{K}}r^{2}}{1+r^{2}} \right) \, \ud F_0(r)
\label{eq:f_Rupperbound}
\\
&= D_{F_0}\exp\left( -\frac{R}{1+\alpha} + \sqrt{{\sf{K}}R}
\right) \label{eq:f_Rupperbound2}
\end{align}
where $0< D_{F_0} \le 1$ is a constant depending on $F_0$. Upper
bound (\ref{eq:f_Rupperbound}) is easily verified by observing
$
\exp\left(- \frac{R}{1+r^2} \right) \le \exp\left(-
\frac{R}{1+\alpha} \right) \,\,\, \forall r \le \sqrt{\alpha}$ and
$I_0 \left( \frac{2\sqrt{{\sf{K}}}r \sqrt{R}}{1 + r^{2}} \right)
\le I_0(\sqrt{{\sf{K}}R}) \le \exp(\sqrt{{\sf{K}}R}) \,\,\,
\forall r \ge 0.
$
Using (\ref{eq:f_Rupperbound2}), we have the following upper
bound:
\begin{align}
\int_0^{\infty} g(R,r) \ln f_R(R;F_0) \, \ud R &\le \ln D_{F_0} -
\frac{1}{1+\alpha}\int_0^{\infty} g(R,r) R \,\, \ud R + \sqrt{K}
\int_0^{\infty} g(R,r) \sqrt{R} \,\, \ud R \\ &\le \ln D_{F_0} -
\frac{1+({\sf{K}}+1)r^2}{1+\alpha} + \sqrt{\sf{K}} \sqrt{1 +
(1+{\sf{K}})r^2} \quad \forall r \ge 0. \label{eq:intupperbound}
\end{align}
The upper bound in (\ref{eq:intupperbound}) follows from the fact
that $g(R,r)$ is a non-central chi-square probability density
function in $R$, and $ \int_0^{\infty} g(R,r) R \,\, \ud R = 1 +
(1+{\sf{K}})r^2$ and $\int_0^{\infty} g(R,r) \sqrt{R} \,\, \ud R
\le \sqrt{1 + (1+{\sf{K}})r^2}$ which follows from the concavity
of $\sqrt{x}$ and the Jensen's inequality. From
(\ref{eq:intupperbound}), we obtain the following upper bound on
the left hand side (LHS) of (\ref{eq:ktcpeak}):
\begin{align}
\text{LHS} \le \ln D_{F_0} - \frac{1+({\sf{K}}+1)r^2}{1+\alpha} +
\sqrt{\sf{K}} \sqrt{1 + (1+{\sf{K}})r^2} + \ln(1+r^2)+ C+1 \quad
\forall r \ge 0. \label{eq:LHSupperbound}
\end{align}
Using the above upper bound, we show that the following assumption
cannot hold true.
\\
i) Assume that the optimal input distribution $F_0$ has an
infinite number of points of increase on a bounded interval. Next
we extend the LHS of (\ref{eq:ktcpeak}) to the complex domain:
\begin{gather}
\Phi(z) = \int_0^{\infty}g(R,z) \ln f_R(R,F_0) \, \ud R +
\ln(1+z^2)+ C+1
\end{gather}
%\vspace{-1cm}
where $z \in \mathbb{C}$. %and $\log$ is the principle branch of the logarithm.
Since the condition in (\ref{eq:ktcpeak}) should be satisfied with
equality at the points of increase of the optimal input
distribution, by the above assumption, $\Phi(z) = 0$ for an
infinite number of points having a limit point. Then by the
Identity Theorem \cite{Knopp}, $\Phi(z) = 0$ in the whole region
where it is analytic. By the Differentiation Lemma \cite [Ch. 12]
{Lang}, one can easily verify that $\Phi(z)$ is analytic in the
region where $\text{Re}(1+z^2) > 0$ which includes the positive
real line. Therefore we conclude that $\Phi(r) = 0 \,\,\, \forall
r \ge 0$. Clearly, this is not possible from the upper bound in
(\ref{eq:LHSupperbound}) which diverges to $-\infty$ as $r \to
\infty$ for any finite $\alpha, {\sf{K}} \ge 0$, and $D_{F_0}>0$.
\\
Reaching a contradiction, we conclude that the optimal
distribution must be discrete with a finite number of mass points.
\hfill $\square$

We note that Theorem \ref{theo:peakRician} establishes the
discrete structure of the optimal input distribution over the
entire {\tsnr} range. The proof basically uses the observation
that a bounded input induces an output probability density
function that decays at least exponentially, and in turn provides
a diverging bound on the Kuhn-Tucker condition. We also note
recent independent work by Huang and Meyn \cite{Meyn}, where the
discrete nature of the optimal input is proven by again showing a
diverging bound on the Kuhn-Tucker condition for a general class
of channels in which the input is subject only to peak amplitude
constraints.

%%%%%%%%%%%%%%%%%%%%%%%%%%%%%%%%%%%%%%%%%%%%%%%%%%%%%%%%%%%%%%%%%%%%%%%%%%%%%%%%%%%%%%%%%
%%%%%%%%%%%%%%%%%%%%%%%%%%%%%%%%%%%%%%%%%%%%%%%%%%%%%%%%%%%%%%%%%%%%%%%%%%%%%%%%%%%%%%%%%
%%%%%%%%%%%%%%%%%%%%%%%%%%%%%%%%%%%%%%%%%%%%%%%%%%%%%%%%%%%%%%%%%%%%%%%%%%%%%%%%%%%%%%%%%
%%%%%%%%%%%%%%%%%%%%%%%%%%%%%%%%%%%%%%%%%%%%%%%%%%%%%%%%%%%%%%%%%%%%%%%%%%%%%%%%%%%%%%%%%

%%%%%%%%%%%%%%%%%%%%%%%%%%%%%%%%%%%%%%%%%%%%%%%%%%%%%%%%%%%%%%%%%%%
%%%%%%%%%%%%%%%%%%%%%%  RICIAN CHANNEL WITH PHASE NOISE  %%%%%%%%%%
%%%%%%%%%%%%%%%%%%%%%%%%%%%%%%%%%%%%%%%%%%%%%%%%%%%%%%%%%%%%%%%%%%%
\vspace{-.3cm}
\section{Rician Fading Channel with Phase Noise} \label{sec:pnRician}

In this section, we deviate from the classical Rician channel
model (\ref{eq:model}) where the specular component is assumed to
be static and  consider the following model \vspace{-.4 cm}
\begin{gather} \label{eq:pnmodel}
y_i = (a_i + me^{j\,\theta_i})x_i + n_i.
\end{gather}
where phase noise is introduced in the specular component. Here,
\{$\theta_i$\} is assumed to be a sequence of independent and
identically distributed uniform random variables on $[-\pi,\pi)$
and $m$ is a deterministic complex constant. We again consider the
noncoherent scenario where $\{a_i\}$ and $\{\theta_i\}$ are known
by neither the receiver nor the transmitter.
%Hence,
%the only difference from the classical Rician model
%(\ref{eq:model}) is that $\theta_i$, is assumed to be uniformly
%distributed on $[-\pi, \pi)$ and not known by the receiver.
This model is relevant in mobile systems where rapid random
changes in the phase of the specular component are not tracked.
Moreover, such a model is suitable in cases where there is
imperfect receiver side information about the fading magnitude. As
another departure from the previous section, here we impose only
an average power constraint, $E\{|x|^2\} \leq P_{av}$. The
discrete nature of the optimal input amplitude follows immediately
from the techniques of Section \ref{sec:Rician} when there is an
additional higher moment constraint.

We immediately realize that the channel output, $y$, is
conditionally Gaussian given $x$ and $\theta$, \vspace{-0.2cm}
\begin{gather} \label{eq:pncondgauss}
f_{y|x,\,\theta}(y|x,\theta) = \frac{1}{\pi(\gamma^2|x|^2+N_0)}
\exp\left(-\frac{\left| y -
me^{j\theta}x\right|^2}{\gamma^2|x|^2+N_0} \right).
\end{gather}
Integrating (\ref{eq:pncondgauss}) over uniform $\theta$, we
obtain the conditional distribution of the channel output given
the input, \vspace{-0.7cm}
\begin{gather} \label{eq:pncond}
f_{y|x}(y|x) = \frac{1}{\pi(\gamma^2|x|^2+N_0)}
\exp\left(-\frac{|y|^2+|m|^2|x|^2}{\gamma^2|x|^2+N_0} \right)
I_0\left(\frac{2|m||y||x|}{\gamma^2|x|^2+N_0} \right).
\end{gather}
We again introduce the following random variables: $R =
\frac{1}{N_0}|y|^2 \,\,,\,\, r = \frac{\gamma}{\sqrt{N_0}}|x|.$
Since the phase information is completely destroyed in the channel
(\ref{eq:pnmodel}) and the above transformations are one-to-one,
we have $ I(x;y) = I(|x|;|y|) = I(r;R). $ Furthermore the
conditional distribution of $R$ given $r$ is easily obtained from
(\ref{eq:pncond}):
\begin{gather}
f_{R|r}(R|r) = \frac{1}{1+r^2}\, \exp\left(-
\frac{R+{\sf{K}}r^2}{1+r^2}\right) I_0\left(
\frac{2\sqrt{{\sf{K}}}r\sqrt{R}}{1+r^2}\right)
\end{gather}
where ${\sf{K}} = \frac{|m|^2}{\gamma^2}$ is the Rician factor.
Similarly as in the previous section, the existence of an optimal
amplitude distribution is shown and the following sufficient and
necessary condition is given in \cite{gursoy}.
\begin{prop:pnKuhn-Tucker}(\underline{Kuhn-Tucker Condition}) \label{prop:pnktc}
For the Rician channel model (\ref{eq:pnmodel}) with an average
power constraint $E\{|x|^2\} \leq P_{av}$, $F_{0}$ is a
capacity-achieving amplitude distribution if and only if there
exists $\lambda \geq 0$ such that the following is satisfied
\begin{align}
-D(f_{R|r}||f_R) + \lambda(r^2-\alpha) + C   &\geq 0 \quad \forall
r \geq 0 \label{eq:pnktc}
%\\ &= 0 \quad \forall r \in E_0
\end{align}
with equality if $r \in E_0$ where $E_0$ is the set of points of
increase of $F_0$.  In the above formulation, $D(\cdot||\cdot)$ is
the divergence (e.g. \cite[Section 2.3]{Cover}), $f_R(R;F_0) =
\int_0^\infty f_{R|r}(R|r) \,\ud F_0(r)$ is the density function
of $R$, $\alpha = \gamma^2\text{\footnotesize{SNR}}$, and $C$ is
the capacity.
\end{prop:pnKuhn-Tucker}
The next theorem gives the main result on the structure of the
optimal input for the Rician fading channel with phase noise.
\begin{theo:pnRician} \label{theo:pnRician}
For the Rician fading channel with uniform phase noise
(\ref{eq:pnmodel}) and average power constraint $E\{|x|^2\} \leq
P$, the capacity-achieving input amplitude distribution is
discrete with a finite number of mass points.
\end{theo:pnRician}
\emph{Proof}: The result is shown by contradiction. Let $F_0$ be
an optimal amplitude distribution. The proof can be summarized as
follows:

i) We first assume that $F_0$ has an infinite number of points of
increase on a bounded interval. The impossibility of this case is
shown by contravening the fact that under this assumption the left
hand side of the Kuhn-Tucker condition which is extended to the
complex domain is identically zero over its region of analyticity.

ii) Then we assume that the optimal distribution is discrete with
an infinite number of mass points but only finitely many of them
on any bounded interval. This assumption is also ruled out by
finding a diverging lower bound on the left hand side of the
Kuhn-Tucker condition.

iii) Having eliminated the above assumptions, we are left with the
only possibility that the optimal distribution is discrete with a
finite number of mass points.

\textbf{Assumption 1}: Assume that $F_0$ has an infinite number of
points of increase on a bounded interval. Then the Kuhn-Tucker
condition (\ref{eq:pnktc}) is satisfied with equality at an
infinite number of points having a limit point.
%We
%will use an approach employed in \cite{Nuriyev} in the context of
%block-independent noncoherent AWGN channels.
First we extend the left hand side of (\ref{eq:pnktc}) to the
complex domain: $ \Psi(z) = -D(f_{R|r=z}||f_R) +
\lambda(z^2-\alpha) + C,$ where $z \in \mathbb{C}$. Equivalently,
using the fact that $f_{R|r}$ is a noncentral chi-square density
function in $R$,
\begin{align} \label{eq:KTCcomplex}
\Psi(z) = \int_0^\infty f_{R|r}(R|z) \ln f_R(R) \,\ud R -&
\int_0^\infty f_{R|r}(R|z)
\ln\left(I_0\left(\frac{2\sqrt{{\sf{K}}}z\sqrt{R}}{1+z^2}\right)\right)
\ud R \nonumber \\ +&\lambda(z^2-\alpha) + \ln(1+z^2) +
\frac{2{\sf{K}}z^2}{1+z^2} + C + 1.
\end{align}
By the Differentiation Lemma \cite[Ch. 12]{Lang}, it is easy to
see that $\Psi$ is analytic in the region where
$\text{Re}\left\{1+z^2\right\} > 0$ and $\text{Re}\{z\}>0$. The
first condition guarantees the uniform convergence of the
integrals in (\ref{eq:KTCcomplex}) by forcing the integrands to
decrease exponentially. Since $I_0$ has zeros on the imaginary
axis and the second integral in (\ref{eq:KTCcomplex}) involves the
logarithm of the Bessel function, with the second condition,
$\text{Re}\{z\}>0$, we exclude the imaginary axis from the region
of analyticity. Since $\Psi(z) = 0$ for an infinite number of
points having a limit point, by the Identity Theorem \cite{Knopp},
$\Psi(z) = 0$ in the whole region where it is analytic. In
particular, we have $\Psi(jb+\frac{1}{n}) = 0$ for all $|b|<1$ and
$n \in \mathbb{Z}^+$, and hence
$
\lim_{n \to \infty} \Psi(jb + \frac{1}{n}) = 0 \quad \forall
|b|<1. %\label{eq:limiting}
$
All the terms other than the second term in (\ref{eq:KTCcomplex})
are analytic also on the imaginary axis with $|z|<1$ and the
limiting expression is obtained by letting $\frac{1}{n} \to 0$ in
the arguments of these functions. For the second term in
(\ref{eq:KTCcomplex}), we need to invoke the Dominated Convergence
Theorem \cite{Rudin} to justify the interchange of limit and
integral. An integrable upper bound on the magnitude of the
integrand of the second integral is shown in \cite[Appendix
D]{gursoy}. Therefore, we have
\begin{align}
\lim_{n \to \infty} \Psi(jb+\frac{1}{n}) =  &\int_0^\infty
f_{R|r}(R|jb) \ln f_R(R) \,\ud R \nonumber \\ -&\int_0^\infty
\lim_{n \to \infty} f_{R|r}\left(R\bigg|jb + \frac{1}{n}\right)
\ln\left(I_0\left(\frac{2\sqrt{{\sf{K}}}\left(jb +
\frac{1}{n}\right)\sqrt{R}}{1+\left(jb +
\frac{1}{n}\right)^2}\right)\right) \ud R \nonumber
\\ +&\lambda(-b^2-\alpha) + \ln(1-b^2) -
\frac{2{\sf{K}}b^2}{1-b^2} + C + 1 = 0 \quad \forall |b| < 1.
\label{eq:pnktcimag}
\end{align}
Note that all the terms other than the second term in
(\ref{eq:pnktcimag}) are real. Next we show that the second term
in (\ref{eq:pnktcimag}) has a nonzero imaginary component yielding
a contradiction. First we evaluate the limit in the integrand as
follows. \vspace{-.1cm}
\begin{align}
&\lim_{n \to \infty} f_{R|r}\left(R\bigg|jb + \frac{1}{n}\right)
\ln\left(I_0\left(\frac{2\sqrt{{\sf{K}}}\left(jb +
\frac{1}{n}\right)\sqrt{R}}{1+\left(jb +
\frac{1}{n}\right)^2}\right)\right) =  \nonumber \\ & = \left\{
\begin{array}{ll}
0 & \text{if } I_0\left( \frac{2\sqrt{{\sf{K}}}jb\sqrt{R}}{1-b^2}
\right) = 0 \\ \frac{1}{1-b^2} \exp \left(-\frac{R -
{\sf{K}}b^2}{1-b^2}\right) I_0\left(
\frac{2\sqrt{{\sf{K}}}jb\sqrt{R}}{1-b^2} \right)
\ln\left(I_0\left( \frac{2\sqrt{{\sf{K}}}jb\sqrt{R}}{1-b^2}
\right)\right) & \text{otherwise},
\end{array} \right. \label{eq:limitingintegrand}
\end{align}
which is obtained easily by observing that all the terms are
analytic in the entire complex plane (excluding $z = \pm j$)
except the logarithm function which is not analytic at the zeros
of the Bessel function. However, as we approach the zeros of the
Bessel function, $I_0 \ln I_0 \to 0$ and hence we obtain
(\ref{eq:limitingintegrand}). Noting that $J_0(z) = I_0(jz)$ where
$J_0$ is the zeroth order Bessel function of the first kind, and
that the set of zeros of the $I_0$ function along the imaginary
axis has measure zero, the second integral in (\ref{eq:pnktcimag})
can now be expressed as follows:
\begin{align}
\int_0^{\infty} \frac{1}{1-b^2} \exp \left(-\frac{R -
{\sf{K}}b^2}{1-b^2}\right) J_0\left(
\frac{2\sqrt{{\sf{K}}}b\sqrt{R}}{1-b^2} \right) \ln\left(J_0\left(
\frac{2\sqrt{{\sf{K}}}b\sqrt{R}}{1-b^2} \right)\right) \ud R.
\end{align}
By applying a change of variables $v =
\frac{2\sqrt{{\sf{K}}}b\sqrt{R}}{1-b^2}$ and expressing $\ln z =
\ln |z| + j \,\text{arg}(z)$, the above integral becomes
\vspace{-.5cm}
\begin{align}
\int_0^\infty &\frac{1-b^2}{2{\sf{K}}b^2} \, v
\exp\left(-\frac{1-b^2}{4{\sf{K}}b^2}v^2 +
\frac{{\sf{K}}b^2}{1-b^2}\right) J_0(v) \ln |J_0(v)| \, \ud v
\nonumber \\ &+ j \pi \int_0^\infty \frac{1-b^2}{2{\sf{K}}b^2} \,
v \exp\left(-\frac{1-b^2}{4{\sf{K}}b^2}v^2 +
\frac{{\sf{K}}b^2}{1-b^2}\right) J_0(v) h(v) \, \ud v.
\label{eq:pnktcimagterm}
\end{align}
In the above formulation $h(v) = 0$ if $v \in (0, \alpha_1)$ and
$h(v) = k$ if $v \in (\alpha_k,\alpha_{k+1})$ where $\{\alpha_k\}$
are the zeros of $J_0$. As noted in \cite{Nuriyev}, the second
term in (\ref{eq:pnktcimagterm}) arises due to the fact that the
logarithm jumps in value by $j \pi$ when a zero is passed. This
fact can be observed by considering the argument of $J_0(b -
j\epsilon)$ as $\epsilon \to 0$ where $b-j\epsilon$ is in a small
neighborhood of $\alpha_k$ such that $J_0(b - j\epsilon)
\backsimeq J_0^{\,\,'}(\alpha_k)(b - j\epsilon - \alpha_k)$.

Using a similar bound obtained in \cite{Nuriyev}, we have
%\vspace{-.1cm}
\begin{align}
&\left|\int_0^\infty \frac{1-b^2}{2{\sf{K}}b^2} \, v
\exp\left(-\frac{1-b^2}{4{\sf{K}}b^2}v^2 \right) J_0(v) h(v) \,
\ud v \right| \ge \nonumber \\ & \ge \left(|J_0(\beta)| - \frac{2
{\sf{K}} b^2}{\beta (1-b^2)}\right)\exp\left(-
\frac{\beta^2(1-b^2)}{4{\sf{K}}b^2} \right) - \left( 1 + \frac{4
{\sf{K}} b^2 }{\pi \alpha_2 (1-b^2)}\right) \exp\left(-
\frac{\alpha_2^2(1-b^2)}{4 {\sf{K}} b^2} \right)
\end{align}

\vspace{-0.2cm} \noindent $\!\!\!$where $\alpha_2$ is the second
smallest zero of $J_0$ on the positive axis and $\beta$ is the
positive zero of $J_1$ less than $\alpha_2$. As noted in
\cite{Nuriyev}, the above lower bound is positive for small enough
values of $b$. In particular, for each ${\sf{K}} > 0$, the above
lower bound is $9.16 \times 10^{-5}$ when $b^2 =
\frac{1}{2{\sf{K}}+1}<1$.

\textbf{Assumption 2}: Next we assume that the optimal
distribution has an infinite number of mass points but only
finitely many of them on any bounded interval. Following an
approach similar to the one used in \cite{Abou}, we first bound
$f_R$ as follows \vspace{-.3cm}
\begin{align}
f_R(R;F_0) = \int_{0}^{\infty} f_{R|r}(R|r) \, \ud F_0(r) &=
\sum_{i = 0}^{\infty} p_i \, f_{R|r}(R|r_i) \\ &\geq p_i \,
f_{R|r}(R|r_i)
\\ &\geq p_i \, \frac{1}{1+r_i^2}
\exp\left(-\frac{R+{\sf{K}}r_i^2}{1+r_i^2}\right) \qquad \forall i
\,\,\, \forall R \geq 0.
\end{align}
where $p_i$ and $r_i$ are the probability and location,
respectively, of the $i^{th}$ mass point of $F_0$. We obtain the
last inequality by using the fact that $I_0(x) \geq 1 \,\,\,
\forall x \geq 0$. Noting that $f_{R|r}(R|r)$ is a noncentral chi
square density function in $R$, this bound leads to the following
lower bound on the left hand side of (\ref{eq:pnktc})
\vspace{-.6cm}
\begin{align}\label{eq:lowerboundonKTC}
\text{LHS} \geq &\ln p_i - \ln (1+r_i^2)-
\frac{1+{\sf{K}}r_i^2}{1+r_i^2} - \frac{({\sf{K}}+1)r^2}{1+r_i^2}
- \int_0^\infty f_{R|r}(R|r)
\ln\left(I_0\left(\frac{2\sqrt{{\sf{K}}}r\sqrt{R}}{1+r^2}\right)\right)
\ud R \nonumber \\ &+\lambda(r^2-\alpha) + \ln(1+r^2) +
\frac{2{\sf{K}}r^2}{1+r^2}+C+1\,\,,\,\, \forall i \,\, \forall r
\geq 0.
\end{align}
\vspace{-.2cm} Noting that
$
I_0\left(\frac{2\sqrt{{\sf{K}}}r\sqrt{R}}{1+r^2}\right) \leq
\exp\left(\frac{2\sqrt{{\sf{K}}}r\sqrt{R}}{1+r^2} \right),
$
we have \vspace{-.3cm}
\begin{align*}
\int_0^\infty f_{R|r}(R|r)
\ln\left(I_0\left(\frac{2\sqrt{{\sf{K}}}r\sqrt{R}}{1+r^2}\right)\right)
\ud R &\leq \frac{2\sqrt{{\sf{K}}}r}{1+r^2} E\{\sqrt{R}|r\}
\\
&\le \frac{2\sqrt{{\sf{K}}}r}{1+r^2} \sqrt{E\{R|r\}}
\\
& = \frac{2\sqrt{{\sf{K}}}r \sqrt{1 + (1 + \K)r^2}}{1+r^2} \le 2
\sqrt{2\K+\K^2}.
\end{align*}
Since we have assumed that the optimal distribution has an
infinite number of mass points with finitely many of them on any
bounded interval, we see that for any $\lambda
> 0$, we can choose an $r_i$ sufficiently large such that the
lower bound (\ref{eq:lowerboundonKTC}) diverges to $\infty$ as $r
\rightarrow \infty$. But by our assumption the left hand side of
(\ref{eq:lowerboundonKTC}) should be equal to zero infinitely
often as $r \rightarrow \infty$, which is a contradiction.
$\lambda = 0$ implies that the power constraint is ineffective.
The impossibility of this case is shown in \cite{Abou}. Therefore
the theorem follows. \hfill $\square$

Recent results \cite{Abou} and \cite{Katz Shamai} have shown the
discrete nature of the optimal distribution for the two special
cases of the model (\ref{eq:pnmodel}): the Rayleigh fading channel
and the noncoherent AWGN channel. We have proven the discreteness
of the capacity-achieving distribution in a unifying setting where
there is both multipath fading and a specular component with
random phase. For the noncoherent AWGN channel, Katz and Shamai
\cite{Katz Shamai} have shown that the optimal input has an
infinite number of mass points. An interesting conclusion of
Theorem \ref{theo:pnRician} is that the presence of an unknown
multipath component induces an optimal distribution with a finite
number mass points. It is also of interest to consider the
classical average-power-limited Rician fading channel
(\ref{eq:model}) for which tight upper and lower bounds on the
capacity were derived in \cite{Lapidoth Moser}. By Proposition
\ref{prop:optphase}, we know that uniform phase is optimal for
this model. Moreover, we have the following partial result on the
optimal amplitude distribution which proves the suboptimality of
input amplitude distributions with unbounded support such as the
Rayleigh distribution.

%%%%%%%%%%%%%%%%%%%%%%%%%%%%%%%%%%%%%%%%%%%%%%%%%%%%%%%%%%%%%%%%%%%
%%%%%%%%%%%%%%%%%%%%%%  RICIAN CHANNEL 22222222222222  %%%%%%%%%%%%
%%%%%%%%%%%%%%%%%%%%%%%%%%%%%%%%%%%%%%%%%%%%%%%%%%%%%%%%%%%%%%%%%%%
\vspace{-.2cm}
\begin{theo:Rician2} \label{theo:Rician2}
For the Rician fading channel (\ref{eq:model}) with only an
average power limitation $E\{|x|^2\} \leq P_{av}$, the optimal
input amplitude distribution has bounded support.
\end{theo:Rician2} \vspace{-.3cm}
\emph{Proof}: Assume $F_0$ is an optimal distribution. We will
prove the proposition by contradiction. So we further assume that
$F_0$ has unbounded support. With this assumption, for any finite
$M \geq 0$, %we obtain the following lower bound: \vspace{-.3cm}                                 %%%%%%%%%%%%%%SV
\begin{align}
f_R(R;F_0) &= \int_{0}^{\infty} g(R,r) \, \ud F_0(r) \nonumber
\\ &\geq \int_{0}^{\infty} \frac{1}{1+r^{2}} \exp \left(
-\frac{R+Kr^{2}}{1+r^{2}} \right) \, \ud F_0(r) \\ &\geq
\int_{M}^{\infty} \frac{1}{1+r^{2}} \exp \left(
-\frac{R+Kr^{2}}{1+r^{2}} \right) \, \ud F_0(r) \\ &\geq
\exp(-\frac{R}{1+M^2}) \int_{M}^{\infty} \frac{1}{1+r^{2}} \exp
\left( -\frac{Kr^{2}}{1+r^{2}} \right) \, \ud F_0(r) \\ &=
D_{F_0,M}\exp(-\frac{R}{1+M^2}) \quad , \quad \forall R \geq 0
\,\,\, \forall M \geq 0 \,\, , \label{eq:lowerboundRician2}
\end{align}
where $0 < D_{F_0,M} \leq 1 \quad \forall M \geq 0$ and $ \forall
F_0 $. Using (\ref{eq:lowerboundRician2}), we obtain the following
lower bound on the left hand side of the Kuhn-Tucker
condition\footnote{The Kuhn-Tucker condition for the
average-power-limited Rician case is essentially the same as
(\ref{eq:ktc}) with $\lambda_2 = 0$.}
\begin{gather}\label{eq:lowerboundRician2int}
\text{LHS} \geq \ln D_{F_0,M} - \frac{1}{1+M^2}
-\frac{1+K}{1+M^2}r^2 +\ln(1+r^2)+\lambda(r^{2}-\alpha)+1+C \,\,\,
, \,\, \forall r\geq 0 \,\,\, \forall M \geq 0. \nonumber
\end{gather}
For any $\lambda > 0$\footnote{The impossibility of $\lambda = 0$
is shown in \cite{Abou}.} and $D_{F_0,M}
> 0$, we can choose $M$ sufficiently large such that the above
lower bound diverges to infinity as $r \rightarrow \infty$.
However, if the optimal input has unbounded support, the LHS of
the Kuhn-Tucker condition should be zero infinitely often as $r
\rightarrow \infty$. This constitutes a contradiction and hence
the theorem follows. \hfill $\square$

%%%%%%%%%%%%%%%%%%%%%%%%%%%%%%%%%%%%%%%%%%%%%%%%%%%%%%%%%%%%%%%%%%%
%%%%%%%%%%%%%%%%%%%%%%   NUMERICAL RESULTS    %%%%%%%%%%%%%%%%%%%%%
%%%%%%%%%%%%%%%%%%%%%%%%%%%%%%%%%%%%%%%%%%%%%%%%%%%%%%%%%%%%%%%%%%%

\section{Numerical Results}  \label{sec:Numerical}

In general, the number of mass points of the optimal discrete
distribution and their locations and probabilities depend on the
{\footnotesize{SNR}}. Analytical expressions for the capacity and
the optimal distribution as a function of {\footnotesize{SNR}} are
unlikely to be feasible. Therefore, we resort to numerical methods
to examine this behavior. The numerical algorithm used here is
similar to the ones employed in \cite{Smith} and \cite{Abou}. In
particular, we start with a sufficiently small
{\footnotesize{SNR}} and maximize the mutual information over the
set of two-mass-point discrete distributions satisfying the input
constraints. Then we test the maximizing two-mass-point discrete
distribution with the Kuhn-Tucker condition. If this distribution
satisfies the necessary and sufficient Kuhn-Tucker condition, then
it is optimal and the mutual information achieved by it is the
capacity. As we increase the {\footnotesize{SNR}}, the required
number of mass points monotonically increases, and therefore to
obtain the optimum distribution we repeat the same procedure for
discrete distributions with increasing numbers of
mass-points.\footnote{We note that the results in this section are
obtained numerically, and hence there is no analytical claim of
optimality.}

\begin{figure}
\begin{center}
\includegraphics[width = 0.7\textwidth]{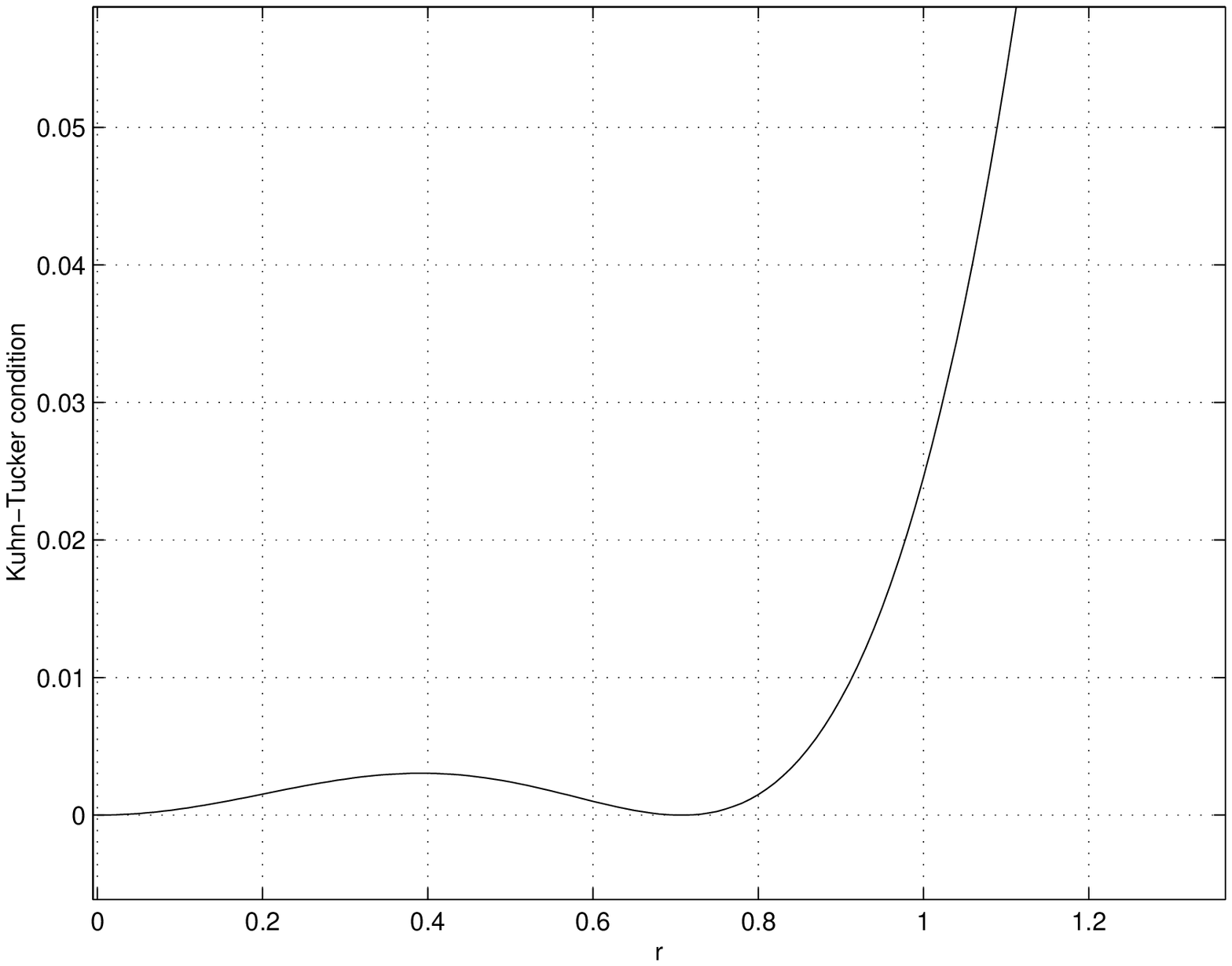}
\caption{The Kuhn-Tucker condition for ${\sf{K}} = 1$, $\alpha =
0.05$ and $\kappa = 10$. $F(r) = 0.9\, u(r) + 0.1 \,
u(r-1/\sqrt{2})$ and $C = 0.0531$. $\lambda_1 = 0.89106$,
$\lambda_2 = 0.15135$ } \label{fig:KTbeta1}
\end{center}
\end{figure}

\begin{figure}
\begin{center}
\includegraphics[width = 0.7\textwidth]{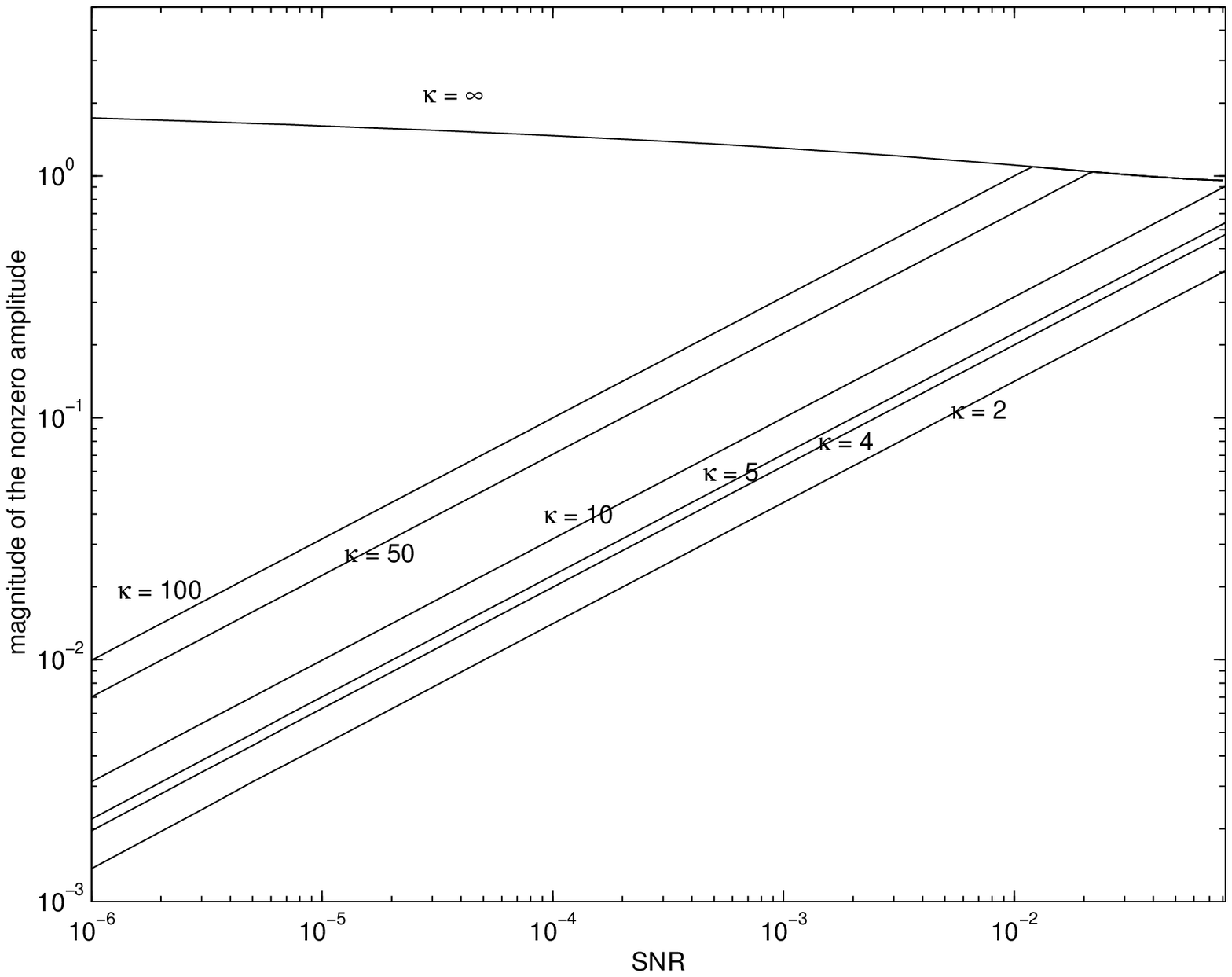}
\caption{Location of the second mass point vs. normalized $\tsnr =
\gamma^2 \frac{P_{av}}{N_0}$ in the Rician channel ${\sf{K}} =
1$.} \label{fig:locationbeta1}
\end{center}
\end{figure}

\begin{figure}
\begin{center}
\includegraphics[width = 0.65\textwidth]{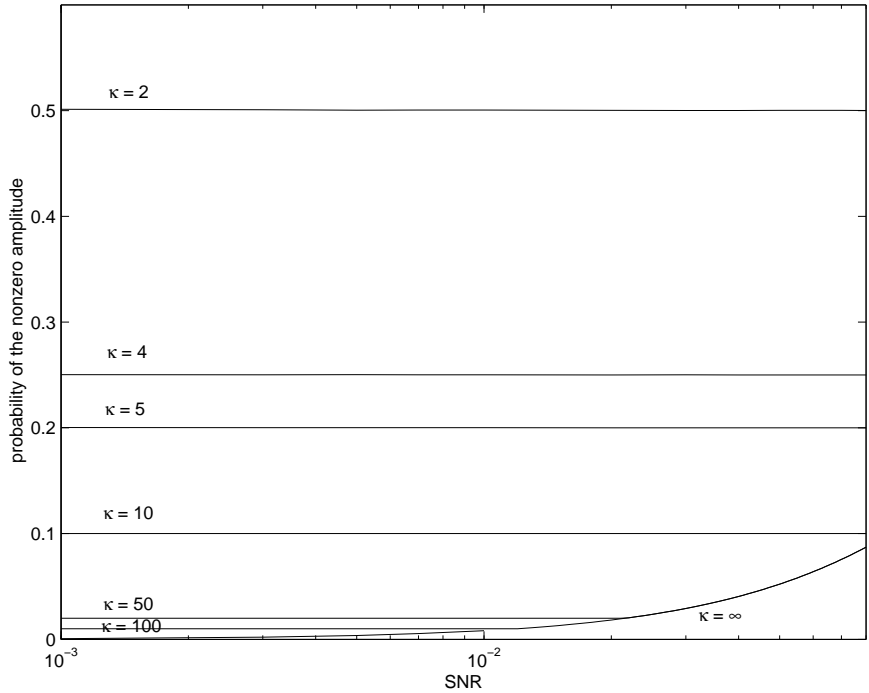}
\caption{Probability of the second mass point vs. normalized
$\tsnr = \gamma^2 \frac{P_{av}}{N_0}$ in the Rician channel
${\sf{K}} = 1$.} \label{fig:probbeta1}
\end{center}
\end{figure}

\begin{figure}
\begin{center}
\includegraphics[width = 0.7\textwidth]{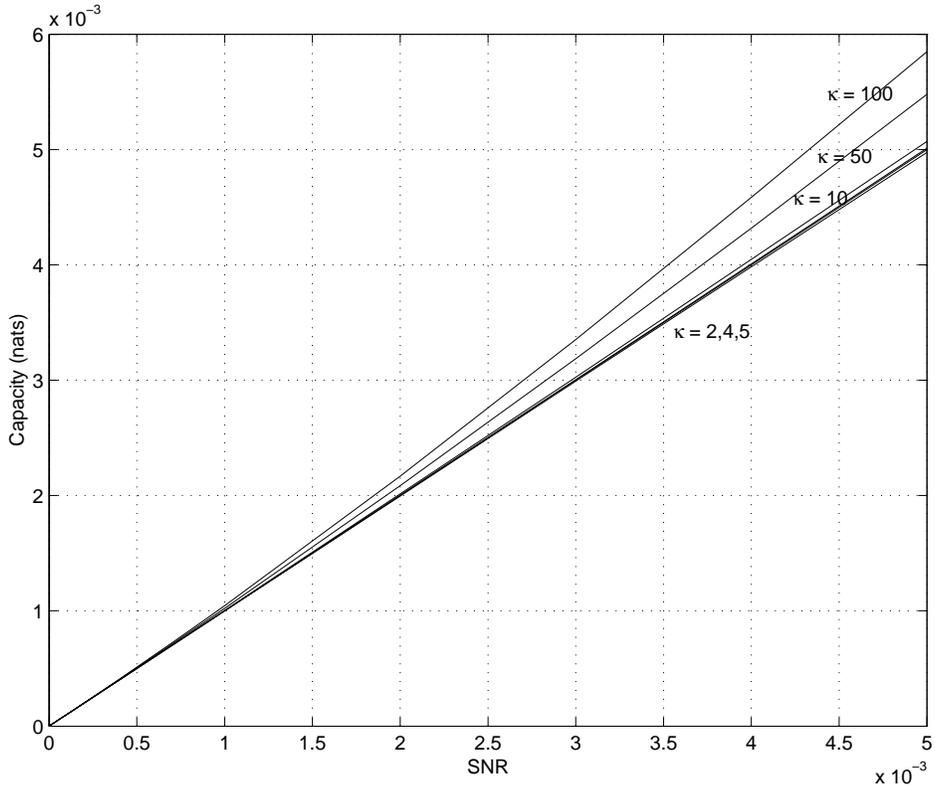}
\caption{Capacity (in nats) vs. normalized $\tsnr = \gamma^2
\frac{P_{av}}{N_0}$ for the Rician channel (${\sf{K}} = 1$)
subject to fourth moment constraints with $\kappa =
2,4,5,10,50,100$.} \label{fig:capacitybeta1}
\end{center}
\end{figure}

For the Rician fading channel ($\K > 0$) with second and fourth
moment input constraints, numerical results indicate that for
sufficiently small {\footnotesize{SNR}} values, the two-mass-point
discrete amplitude distribution \vspace{-.5cm}
\begin{gather} \label{eq:optinput}
F(|x|) = \left(1-\frac{1}{\kappa} \right) u(|x|) +
\frac{1}{\kappa} u(|x|-\sqrt{\kappa N_0 \ssnr})
\end{gather}
is optimal. Note that this distribution does not depend on the
Rician factor ${\sf{K}}$. Figure \ref{fig:KTbeta1} plots the left
hand side of the Kuhn-Tucker condition (\ref{eq:ktc}) as a
function of $r$ for the distribution $F(r) = 0.9 u(r) + 0.1 u(r -
1/\sqrt{2})$ for the Rician fading channel (${\sf{K}} = 1$) with
$\alpha = 0.05$ and $\kappa = 10$. From the figure we see that the
Kuhn-Tucker condition is satisfied and the optimal distribution is
in the form given by (\ref{eq:optinput}). Figures
\ref{fig:locationbeta1} and \ref{fig:probbeta1} plot the magnitude
and the probability of the nonzero amplitude respectively as a
function of {\footnotesize{SNR}} ($N_0 = 1$) for various values of
$\kappa$. We immediately notice the significant impact of imposing
a fourth moment constraint. When there is no such constraint, the
nonzero amplitude migrates away from the origin as
$\textrm{\footnotesize{SNR}} \rightarrow 0$ while its probability
decreases sufficiently fast to satisfy the average power
constraint. This type of input is called \emph{flash signaling} in
\cite{Verdu}. However, as we see from Figures
\ref{fig:locationbeta1} and \ref{fig:probbeta1}, if there is a
fourth moment constraint with a finite $\kappa$, then the behavior
is quite different. The nonzero amplitude approaches the origin as
$\textrm{\footnotesize{SNR}} \rightarrow 0$ while its probability
is kept constant. In the Rayleigh channel ($\K = 0$),
(\ref{eq:optinput}) is still optimal at low $\tsnr$ up to a point
after which, as $\tsnr$ is further lowered, the second moment
constraint becomes inactive and we observe that the nonzero mass
point approaches the origin more slowly while its probability
decreases. From Fig. \ref{fig:capacitybeta1}, which plots the
capacity curves as a function {\footnotesize{SNR}} for various
values of $\kappa$ in the low-power regime, we see that all the
curves have the same first derivative at zero {\tsnr}. This may
suggest that performance in the low-power regime is similar for
any finite value of $\kappa$. However, as we shall see in
\cite{part2}, the picture radically changes when we investigate
the spectral-efficiency/bit-energy tradeoff.

\begin{figure}
\begin{center}
\includegraphics[width = 0.7\textwidth]{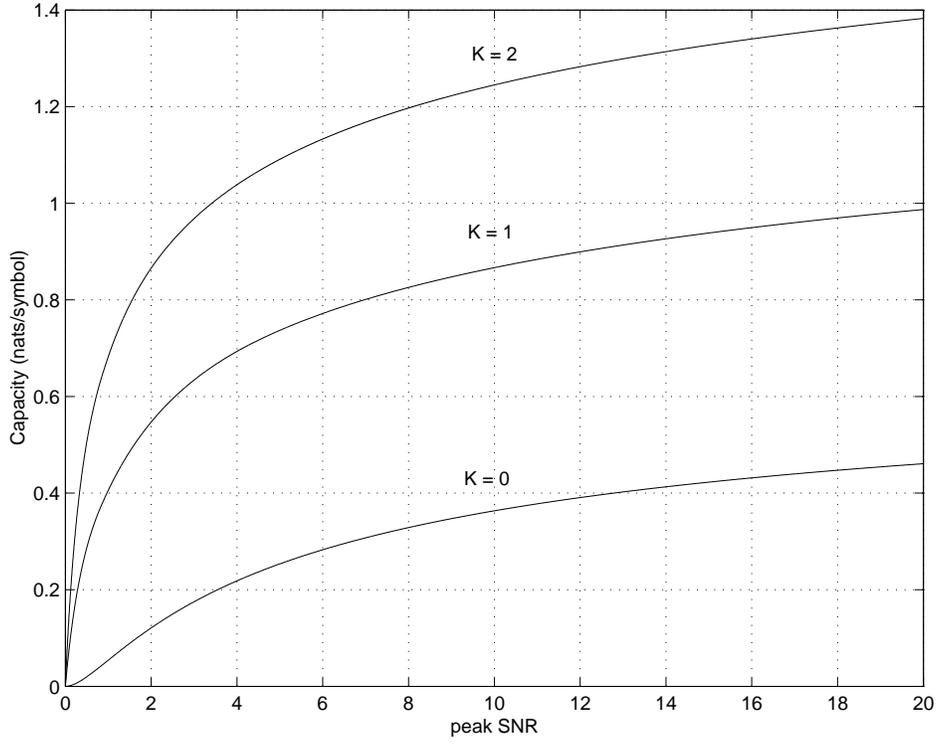}
\caption{Capacity curves as a function of the normalized peak
$\tsnr = \gamma^2 \frac{P}{N_0}$ for the peak-power limited
Rayleigh channel ($\K = 0$) and Rician channels with ${\sf{K}} =
1,2$.} \label{fig:capK012}
\end{center}
\end{figure}

For the peak-power limited Rician fading channel ($\K > 0$),
numerical results indicate that for sufficiently low {\tsnr}
values, the optimal amplitude distribution has a single mass at
the peak level $\sqrt{P}$ and hence all the information is carried
on the uniform phase. For the Rayleigh channel ($\K = 0$), an
equiprobable two-mass-point distribution where one mass is at the
origin and the other mass at the peak level is capacity-achieving
in the low-power regime. Fig. \ref{fig:capK012} plots the capacity
curves for the peak-power limited Rayleigh channel and Rician
channels with $\K = 1,2$ as a function of the peak
{\scriptsize{SNR}}. Note that the Rayleigh channel capacity curve
has a zero slope at zero {\scriptsize{SNR}}.

\begin{figure}
\begin{center}
\includegraphics[width = 0.7\textwidth]{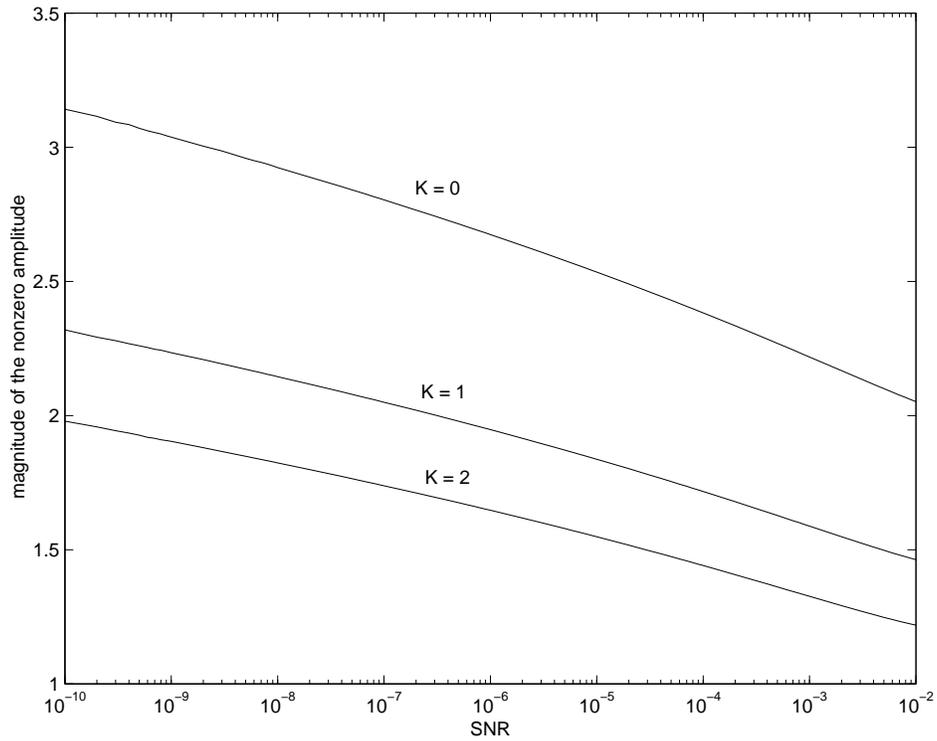}
\caption{Magnitude of the nonzero amplitude vs. normalized $\tsnr
= \gamma^2 \frac{P_{av}}{N_0}$ for the Rician channel with phase
noise ${\sf{K}} = 0,1,2$.} \label{fig:pnlocation}
\end{center}
\end{figure}

\begin{figure}
\begin{center}
\includegraphics[width = 0.65\textwidth]{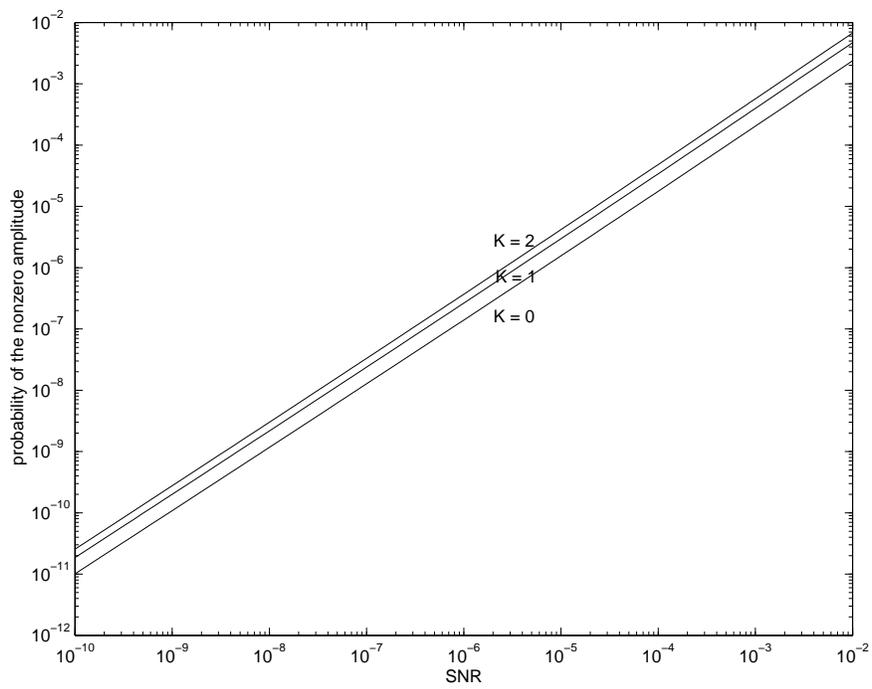}
\caption{Probability of the nonzero amplitude vs. normalized
$\tsnr = \gamma^2 \frac{P_{av}}{N_0}$ for the Rician channel with
phase noise ${\sf{K}} = 0,1,2$.} \label{fig:pnprob}
\end{center}
\end{figure}

\begin{figure}
\begin{center}
\includegraphics[width = 0.7\textwidth]{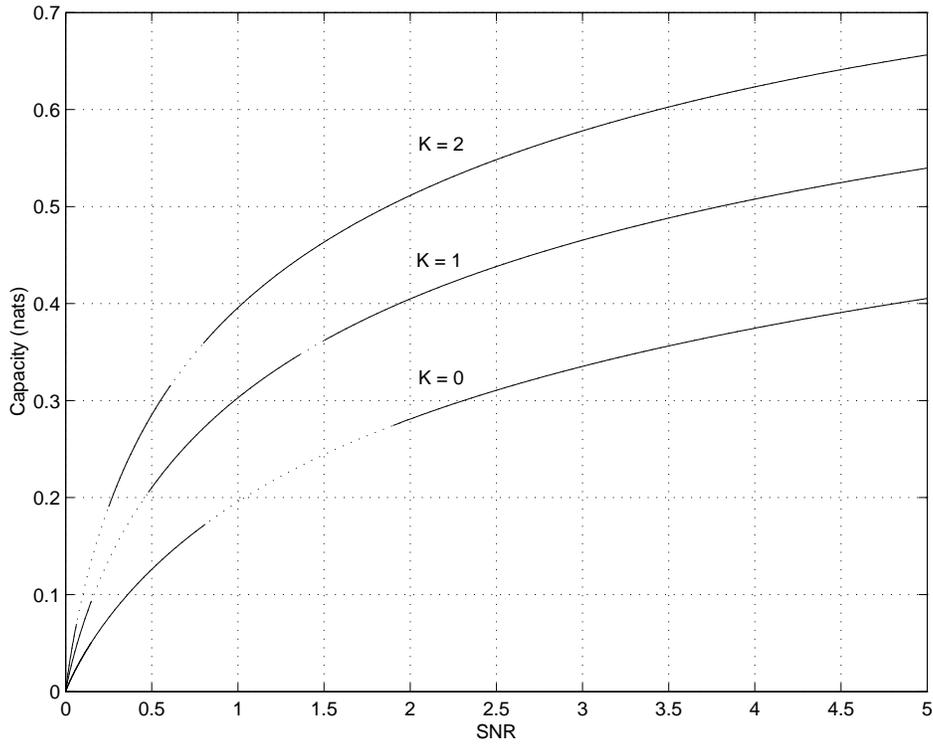}
\caption{Capacity curves as a function of the normalized $\tsnr =
\gamma^2 \frac{P_{av}}{N_0}$ for the Rician channel with phase
noise with Rician factors ${\sf{K}} = 0,1,2$. The dashed segments
are interpolated capacity curves. Numerical optimization methods
do not provide stable results in these regions where a new mass
point is emerging with a very small probability. }
\label{fig:pncapacity}
\end{center}
\end{figure}

For the Rician fading channel with phase noise (\ref{eq:pnmodel}),
numerical results illustrate again that a two-mass-point discrete
distribution is optimal for sufficiently small
{\footnotesize{SNR}} values. Figures \ref{fig:pnlocation} and
\ref{fig:pnprob} plot the magnitude and probability, respectively,
of the optimal nonzero amplitude for this channel with Rician
factors ${\sf{K}} = 0,1,2$. Note that only an average power
constraint is imposed here. We observe that flash-signaling-type
optimal input, where the nonzero amplitude migrates away from the
origin as $\text{\footnotesize{SNR}} \rightarrow 0$ while its
probability is decreasing, is required in the low-power regime.
For fixed {\footnotesize{SNR}}, we also see that the nonzero
amplitude is closer to the origin for higher Rician factors
{\sf{K}}. Finally Fig. \ref{fig:pncapacity} provides the capacity
curves as a function of {\footnotesize{SNR}} for Rician factors
${\sf{K}} = 0,1,2$.

\section{Conclusion} \label{sec:conclusion}

In this paper, we have analyzed the structure of the
capacity-achieving input for the noncoherent Rician fading
channel. We have limited the peakiness of the input by imposing a
fourth moment or a peak constraint. Using a sufficient and
necessary condition, we have proven that when the input is subject
to second and fourth moment limitations, the optimal input
amplitude is discrete with a finite number of levels in the
low-power regime. It turns out that a particular two-mass point
distribution that depends only on the {\tsnr} and $\kappa$ is
asymptotically optimal as $\tsnr \rightarrow 0$. Discreteness of
the optimal input amplitude distribution has also been shown for
the peak-power limited Rician channel over the entire {\tsnr}
range. This time, the amplitude distribution with a single mass at
the peak level is optimal in the low-power regime for the Rician
channel with $\K > 0$.

We also have analyzed a Rician fading channel model where there is
phase noise in the specular component. We have shown that under an
average power limitation, the optimal amplitude is discrete with a
finite number of levels. For this model, we have provided
numerical results for the capacity and the optimal input
distribution where we observed that a flash-signaling-type input
is required in the low-power regime. We have also proved that the
optimal input for the average-power-limited classical Rician
channel has bounded support.

\end{spacing}
\begin{spacing}{0.8}

\newpage

\end{spacing}
\end{document}